\documentclass{article}
\usepackage[hmargin=.75in,vmargin=1in]{geometry}
\usepackage[american]{babel}
\usepackage[T1]{fontenc}
\usepackage{times}

\usepackage{epsfig,graphicx}
\usepackage{amsfonts,amsmath,amssymb}
\usepackage[hyphens]{url}
\usepackage{hyperref}

\title{\bf  How to transform the Apple's application 'Find My' into a toolbox for whistleblowers}           

\author{
{\bfseries Amadou Moctar Kane}\\
KSecurity  Dakar Senegal\\
amadou1@gmail.com
}

\begin{document}

\maketitle



\begin {abstract}
The recent introduction of Find My app by Apple will open a large window of opportunities for whistleblowers.
Based on a short range Bluetooth signals, an EC P-224 encryption, and an end-to-end encrypted manner using iCloud Keychain, Find My app is probably the first application broadcasting a large number of anonymous public key on this scale.

\noindent Hence, this new Apple's application may introduce a revolution in secret communication, if we divert it from its primordial use and transform it into a powerful tool to put in the hands of whistleblowers.

\noindent By using Find My app and an entity authentication protocol based on artificial intelligence, our goal is to make mass surveillance and kleptographic backdoors ineffective in the lifting of the whistleblower's anonymity.

\noindent However, in some case, Find my app may also be a powerful tool in the hands of dictatorships governments in their fight against whistleblowers and political adversaries.

\noindent Thus, the aim of this paper is to show with simple examples, how these two previous situation can happen.

\end {abstract}

\medskip
\section{ Introduction}
Introduced in 2019 by Apple, the Find My app can help users to locate their device even if that latter is offline. Find My app is in iOS, iPadOS, and macOS. 
Before that, an online (WiFi, cellular network) device was able to report its location to the owner via iCloud, while an offline device was no able to send its location. 
Find My app is designed for offline devices by sending out short range Bluetooth signals from the missing device that can be detected by other Apple devices in use nearby. 
It works even if the device is offline and asleep, however the use of Find My app is optional. When a device has offline finding enabled, it also means that it can be located by other participants in the same way. 
Those nearby devices will send the detected location of the missing device to iCloud allowing the owner to locate it in Find My app.
Apple certify that it protect the privacy and security of all the users involved in the previous process.

\subsection{End-to-end encryption in Find My}

As presented in \cite{Apple2019}, Find My app is built on a foundation of public key cryptography. When offline finding is enabled in Find My settings, an EC P-224 private encryption key pair noted ${d,P}$ is generated directly on the device where $d$ is the private key and $ P$ is the public key. Additionally, a 256-bit secret $SK_0$ and a counter $i $ is initialized to zero. This private key pair and the secret are never sent to Apple and are synced only among the user's other devices in an end-to-end encrypted manner using iCloud Keychain.

The secret and the counter are used to derive the current symmetric key $ SK_i $with the following recursive construction: $SK_i= KDF(SK_{i-1},“update”)$.

Here, the $KDF$ seems to be the Concatenation Key Derivation Function (Approved Alternative 1) as described in 5.8.1 of NIST SP 800-56A. 

Based on the key $SK_i$ two large integers $ u_i$ and $ v_i$ are computed with $ (u_i,v_i) = KDF(SK_i, “diversify”)$.

Both the P-224 private key denoted $ d$ and corresponding public key referred to as $P$ are then derived using an affine relation involving the two integers to compute a short lived key pair: the derived private key is $ d_i $ where $d_i= u_i* d + v_i $ (modulo the order of the P-224 curve) and the corresponding public part is $P_i$ and verifies $P_i= u_i*P + v_i*G$.
When a device goes missing and can not connect to WiFi or cellular - for  example, a MacBook left on a park bench - it begins periodically broadcasting the derived public key $ P_i$ for a limited period of time in a Bluetooth payload. By using P-224, the public key representation can fit into a single Bluetooth payload. The surrounding devices can then help in the finding of the offline device by encrypting their location with the public key. 
Approximately every 15 minutes, the public key is replaced by a new one using an incremented value of the counter and the process above so that the user can not be tracked by a persistent identifier. The derivation mechanism prevents the various public keys $P_i$ from being linked to the same device.

\medskip

{\bf Locating missing devices in Find My }

\medskip

Any Apple devices within Bluetooth range that have offline finding enabled can detect this signal and read the current broadcast key $P_i$. Using an ECIES (Elliptic Curve Integrated Encryption Scheme) construction and the public key $P_i$, the finder devices encrypt their current location information and relay it to Apple. The encrypted location is associated with a server index which is computed as the SHA-256 hash of the P-224 public key $ P_i$ obtained from the Bluetooth payload.

The owner of the missing device can reconstruct the index and decrypt the encrypted location, more information on this process can be found in \cite{Apple2019}.

When offline finding is disabled, the device no longer acts as a finder nor is it detectable by other finder devices. 
The traffic sent to Apple by finder devices contains no authentication information in the contents or headers.

\subsection{Whistleblower}

The whistleblower is a person who exposes any kind of information or activity that is deemed illegal, or unethical within an organization and his adversaries are often very powerful.

There  already exists some tools which have been developed for whistleblowers, however using them seems to be very difficult. For example, if the whistleblower Alice is working for a corrupted government, she has to find and to talk with journalists or activists online, which is a very big risk, since most of journalists or activists can be under surveillance. 

Other solutions, such as the use of TOR \cite{Dingledine} can create suspicion and surveillance for whistleblowers, due to the fact that she can be under a mass surveillance program.

Hence, we may assume that the mass surveillance technologies can alert the whistleblower's supervisor if she installs Tor or other software of this type on his personal computer, they can also store and scrutinize all her communication on the Internet.

It can also be noted that TOR has other worries, since a lot of websites present in the Darkweb - which were using it - had been found and closed.

Finally, at the era of mass surveillance \cite{Gellman}, it makes no sense to think that with a small number of relays and bridges \cite{Metrics}, TOR can defeat the tracking of whistleblowers. We can remember that the means available to organizations practicing mass surveillance are colossal.

For the security of whistleblowers be assured there, TOR should have a number of nodes that mass surveillance agencies will not be able to monitor. The other problem is due to the fact that becoming a TOR node means bringing support to pedophils, spy and traffickers who are today the main beneficiaries of that network.

It can also be noted that, there are some whistleblowers who use Virtual Private Network (VPN), however, the security of this latter depends on the probity of its owner and his links with some governments.

Apart from TOR and VPN, other tools have been proposed to help whistleblowers in their information dissemination, hence in \cite{Moctar}, the whistleblower will send the encrypted confidential information to all his contacts on social networks, these latter would do so until this information arrives to the person who had received the key from the plants or animals DNA. 

However, this solution has two problems, the fact that a whistleblower sends an encrypted message can be an offense in some countries, and may arouse suspicions in others. Similarly deposing the DNA on the street can be dangerous for a whistleblower who is under surveillance.

Later, Richter et al. \cite{Richter} have proposed ways to defeat arbitrary tracking in printed papers, this type of method should be coupled with any solution designed for the protection of whistleblowers, including the methods that we present in this paper. Thus, the first action of the whistleblower should be to defeat watermark inserts in the documents seized. We may think about a software that would read the confidential information and rewrite it in a new file created for this purpose and free of watermark. 

Similarly, it is important that the whistleblower erases or encrypts all the confidential informations presents in his device, just after their transmission, to avoid any form of forensic investigation.

\subsection{Background}

\subsubsection{Bluetooth}

The use of Bluetooth does not need an Internet Provider, which depend on government's licences to operate.  This licence may oblige these companies to participate on a targeted surveillance due to the fact that some governments may threaten them. In addition to that, theses networks are in the heart of mass surveillance (Internet, cellular etc.). 
That is the reason why, in this paper we will avoid the use of the Internet and cellular network as far as possible.
As presented in \cite{WikipediaBluetooth}, the Bluetooth is a wireless technology standard used for exchanging data between fixed and mobile devices over short distances using short-wavelength UHF radio waves in the industrial, scientific and medical radio bands, from 2.400 to 2.485 GHz, and building personal area networks (PANs).
Bluetooth exists in numerous products such as telephones, speakers, tablets, media players, robotics systems, laptops, and console gaming equipment as well as some high definition headsets, modems, hearing aids and watches.

As claimed by Apple \cite{Apple2019} for Find My app, we also suppose here that the Bluetooth communication  is encrypted, anonymous, and designed to be battery and data efficient, so there is minimal impact on battery life cellular data plan usage and user privacy is protected. We also assume that a proximity is determined through Bluetooth Low Energy (BLE) advertisements.
In this paper, the encrypted confidential information will be broadcasted such as the broadcast of Apple for the derived public key $P_i$.

As presented in Gunnar's comparison, Bluetooth Low Energy ( BLE) range may be  up to 400m indoor and 1000m outdoor \cite{Sponas}.  

However in \cite{Celosia}, Celosia et al. showed that, in Apple's devices, if BLE was designed to provide a range of up to 100 meters in outdoor environments, it is usually shorter in practice. They measured the BLE range of several Apple's devices and found that they can all be received at least to 61 and 38 meters respectively in outdoor and indoor environments.
In this study, we will suppose that the whistleblower can capture and generate the bluetooth packets which are  used by Find My app. 
Which is already the case for  Apple Continuity messages in many studies, for example, in \cite{Celosia},  capture and generation of advertisement packets were respectively done using a sniffer based on the bluepy python library and a bash script based on hcitool. In both cases, they used a CSR v4.0 Bluetooth USB dongle as transceiver.

\subsection{Kleptographic Backdoor}

As defined in wikipedia, a backdoor is a typically covert method of bypassing normal authenication or encryption in a computer, product, \dots.

A kleptographic attack \cite{WikipediaKleptography} is an attack which uses asymmetric cryptography to implement a cryptographic backdoor.

For example, the Dual\_EC\_DRBG cryptographic pseudo-random number generator from the NIST SP 800-90A is thought to contain a kleptographic backdoor. Dual\_EC\_DRBG utilizes elliptic curve cryptography, and NSA is thought to hold a private key which, together with bias flaws in Dual\_EC\_DRBG, allows NSA to decrypt SSL traffic between computers using Dual\_EC\_DRBG for example.

 Hence, in our models, we will try to defeat this type of kleptographic backdoor in the context of mass surveillance where the attacker ignore the owner of the public key, while he has a private key allowing him to decrypt any message encrypted with a public key $P_i$ obtained from Find My app.

\subsection {Continued Fraction}
\medskip
{\bf Continued Fractions :} An expression of the form 
$$
\begin{array}{lll}
\alpha & =&  a_0 + \displaystyle \frac{{b}_0}{\displaystyle a_1
           + \frac{{b}_1}{\displaystyle a_2
             + \frac{{b}_2}{\ddots}}}  \\
        \end{array}
$$
is called a generalized continued fraction. Typically, the numbers $a_1, \dots, b_1, \dots$  may be real or complex, and the expansion may be finite or infinite.

\noindent We will avoid the use of the continued fraction expansions involving ${b}_i= 1$ for most $i$'s. However, in order to simplify our explanation we will use in some cases the classical continued fraction expansion, namely ${b}_i=1$ for any $i$:
$$
\begin{array}{lll}
\alpha & =&  a_0 + \displaystyle \frac{1}
            {\displaystyle a_1
           + \frac{1}{\displaystyle a_2
             + \frac{1}{\ddots}}}  \\
        &=:&  [a_0,a_1,a_2,a_3, a_4, \ldots].
\end{array}
$$

\noindent In this paper we denote by $\Gamma$ the combined sets of algebraic irrationals of degree greater than 2 and transcendental numbers. Our algorithm, will use the irrational numbers which are in $\Gamma$, but we will avoid the use of transcendental numbers having a predictable continued fraction expansion (some examples of irrational numbers given a predictable continued fraction expansion are presented in \cite{{Beeler},{Knuth}}).

\noindent To calculate the classical continued fraction expansion of a number $\alpha$, write down the integer part of $\alpha$. Subtract this integer part from $\alpha$. If the difference is equal to 0, stop; otherwise find the reciprocal of the difference and repeat. The procedure will halt if and only if $\alpha$ is rational.

\noindent We can enumerate some continued fraction properties:
\begin{enumerate}
	\item The continued fraction expansion of a number is finite if and only if the number is rational.
	\item The continued fraction expansion of an irrational number is unique.
	\item Any positive quadratic irrational number $\alpha$ has a continued fraction which is periodic from some point onward, namely a sequence of integers repeats. (Lagrange Theorem)
	\item The knowledge of the continued fraction expansions of $\alpha$ and $\beta$ cannot determine simply that of $\alpha + \beta$, or $\alpha\beta$.
\end{enumerate}

\noindent{\bf Notation}
Let $\alpha \in \Gamma$ and $[{a}_{1},\dots,{a}_{m},\dots,{a}_{m+n},\dots]$ be the continued fraction expansion of $\alpha$; $m$ and $n$ are two integers such that $m >1$ and $n\geq 1$. We denote $\delta$ the vector made with the $n$ partial quotients following the $m$ first partial quotients in the continued fraction expansion.

\medskip
\noindent \textbf{Result 1.}
{\sl It is not possible to find $\alpha$ out of the knowledge of $\delta$.}

\medskip
\noindent {\sc Proof.} Let ${\alpha}\in \Gamma$. We suppose that we know a given part $[{a}_{m+1}, \dots,{a}_{m+n}]$ of ${\alpha}$'s continued fraction expansion. Can we find ${\alpha}$ with the knowledge of these $n$ partial quotients?
The answer is negative, because there exist an infinite number of irrationals with these same partial quotients.
For instance we can exhibit infinitely many irrational numbers $\alpha_{\rho}$ which are different from $\alpha$ and which have the property that $[{a}_{m+1}, \dots,{a}_{m+n}]$ appears as a sequence of $n$ consecutive partial quotients. As a matter of fact, when $\theta$ is an irrational number, it suffices to consider any sequence of $m$ integers ($r_1,r_2,\dots, r_m$) and to define $\alpha_\rho$ to be  
$$
\begin{array}{lll}
\alpha_\rho = r_1 + \displaystyle \frac{b}
            {\ddots \displaystyle r_{m}
             + \frac{ b}{\displaystyle a_{m+1}
             + \frac{b}{\ddots \displaystyle a_{m+n-1}
             + \frac{ b}{\displaystyle a_{m+n}
               + \frac{b}{\displaystyle \theta }}}}}  \\
            =: [r_1, r_2, r_3,\dots, r_{m}, a_{m+1}, \dots, a_{m+n},\theta].
\end{array}
$$

\textbf{Result 2:} For an integer $r$ such that  $r\geq 3 $ and a real algebraic number $A$  $ (A > 1)$, the number  $\sqrt[r]{\log(A)}$  is transcendental.

\textbf{Proof .} Assume that $A$ is a real algebraic number such that $A > 1$, then  $\log(A)$ is transcendental number by Corollary 3.6 of \cite{Burger}.

If we suppose that   $X=\sqrt[r]{\log(A)}$ is an algebraic number, then  $X^r$ is a algebraic number, which is absurd because $X^r=\log(A)$ and $\log(A)$  is transcendental.

Thus, in view of the previous result, the transcendental number which we will use in our authentication protocol will be in this form $ \sqrt[r]{\log(N_a)}$, where $ N_a$ is the nonce chosen randomly by the principal.

\subsection {Our contribution}
\subsubsection{Advantages}
By diverting Find My app, initially destined to find lost or stolen devices, to another end, we will use the models of this section to allow the whistleblowers to broadcast confidential documents, while protecting their anonymity.
 These tools could need the collaboration of their usual partners which are journalists and activists.
We also considered in this article that, in some cases, the whistleblower may refuse to trust the journalists preferring to transmit this information to the people directly, it will belong to the population to disseminate it massively in order to stop the negative actions of governments.
The whistleblower has several possibilities permitting him to disseminate the confidential information with  Find My app, however,  here, we will limit ourself to four sample examples which may cover many other cases. Depending on the possibilities of the whistleblower,  other models may be found due to the fact that Find My app has open a boulevard of opportunities.
\begin{enumerate}
\item The case where the whistleblower knew the journalist or the activist who have to disseminate the information before the beginning of the process.
\item The case where the whistleblower did not know the journalists or activists before the start of the operation, however, he will discover them during the operation.
\item The case where the whisltleblower does not trust the journalists and activists, and prefers to broadcast the confidential information directly to the population. 
\item The case where the whistleblower knows the location and time of the meeting but he ignores the identity of the journalist. 
\end{enumerate}

\subsubsection{Disadvantages}
By using the elements made available to the general public and users of Find My app, we also found some situations which can be hazardous for whistleblowers and vulnerable people. 
We have in these schemes two types of opponents:

\begin{itemize}
	\item   A powerful opponent who can cut the Internet and Cellular network throughout a city \cite{Shutdown}. This opponent is often a state, cyber criminal working on behalf of a state, or for an criminal organization. 
	\item An opponent weaker than the previous, but who may lead passive attacks using the Apple tools (iPhone ...), compatible Bluetooth sensors, interception software \dots

\end{itemize}

This paper is organized as follows, In section two, we will present the benefits that the Find My application could bring to whistleblowers. Before the conclusion, we will detail situations where the Find My app would endanger the whistleblowers and vulnerable people.

\section{Advantages that Find My app may bring to whistleblowers}
\subsection{notation}
\begin{itemize}
	\item In these schemes, Alice will be the whistleblower, Bob will be the journalist or the activist. The secret services (SS) will designate the services responsible for the fight against whistleblowers. Tartuffe is the main adversary of Alice, outside his secret services, he has several thousand members of his klan who may help him in his fight against journalists and whistleblowers, we will note the members of the clan Pongo.
           \item $||$ will means a separation between two parts of the message.
      \item $|$ will designate a concatenation between two public keys.
\item Anonymous users are people who want to help whistleblowers, activists, and journalists to fight  bad governance, however they do not want to be in front of the scene and prefer for example to carry encrypted information.
\end{itemize}
	
\subsection{Definition}

As defined by Apple , we use in this paper the same EC P-224 private encryption key pair noted $ {d_i,P_i} $ generated on the device, where $d_i$ is the private key and $P_i$ is the public key. We recall that $P_i$ is broadcast for a limited period of time in a Bluetooth payload.  
\begin{itemize}
	\item We denote by $E(X:Y)$, the cipher obtained from a public key encryption such that $X$ is the plaintext, and $Y$ the public key.
\item We denote by ${P_i}_{user}$ the public key generated by the user's device.
\item The hash function $H$ used here is SHA-256.
\end{itemize}

\subsection{Authentication protocol based on Artificial Intelligence}

In order to protect whistleblowers and journalists, we will use in some cases an authentication algorithm which protect the anonymity of its users.
Thus, the best way to preserve their anonymities is to remove the identity of the principal in the authentication processes.

\subsubsection{The Needham-Schroeder Protocol}

As defined in \cite{Needham} the public key protocol consists on the following seven steps:

\medskip

Step 1:	$ A \rightarrow   AS $

\medskip

The exchange opens with A consulting the authentication server (AS) to find B's public Key.

\medskip

Step 2: AS responds with: $E(PK_B || B : SK_{AS})$

\medskip

Where $SK_{AS}$ is the authentication server's secret key,  $PK_B$ is B's public key and B is B's identity.

\medskip

Step 3: A sends to B the following $E(N_a || A : PK_B)$

\medskip

This step is for the communication with B to be initiated. This message, which can only be understood by B indicates that someone purporting to be A wishes to establish  communication with B. B decrypts the message with his private key and then finds the nonce $N_a$  chosen by A.

\medskip

Step 4 \& 5: B finds A's public key $PK_A$ with steps similar to 1 \& 2.

\medskip

Step 6: At this point B return the nonce $N_a$, along with a new nonce $N_b$ , to A, encrypted with A's public key $E(N_a || N_b : PK_A)$.

\medskip

Step 7: At the end, A returns the nonce $N_b$  to B, encrypted with B's public key. 

\medskip

	\textbf{The protocol can be described as follows:}

\medskip

1 $ A 	 \rightarrow   AS	$:  	$A  \, B$

\medskip

2 $ AS 	  \rightarrow A       $ 	:   	 $E(PK_B || B : SK_{AS})$

\medskip

3.$ A   \rightarrow B $    	:    	$E(N_a || A : PK_B)$

\medskip

4.$ B  \rightarrow    AS	$:   	$B     \, A$

\medskip

5.$ AS 	\rightarrow B $ 	:   	$E(PK_A || A: SK_{AS})$

\medskip

6.$ B 	\rightarrow A $ 	:    	 $E(N_a || N_b : PK_A)$

\medskip

7.$ A   	\rightarrow B $ 	:     	 $E( N_b : PK_B)$

\medskip

\subsubsection{The Needham-Schroeder-Lowe Protocol}

In \cite{Lowe} Lowe shows that an attack on the protocol may allow an Tartuffe's secret service member named SS to impersonate the whistleblower A to set up a false session with the journalist B. 

In this attack, we can ignore the interaction with the server because this does not have a real influence on this attack. The attack involves two simultaneous runs of the protocol: in run 1, A establishes a valid session with SS, and in run 2, SS impersonates A to establish a fake session with B. 

In step 1.3, A starts to establish a normal session with SS, sending him a nonce $N_a$. 

In step 2.3, the intruder impersonates A by trying to establish a false session with B, sending him the nonce  obtained in the previous message. 

B responds in the message 2.6 by selecting a new nonce $N_b$ , and trying to return it along with $N_a$ , to A. The secret service member therefore forwards the message to A in the step 1.6.  A decrypts the message to obtain $N_b$ , and returns this to SS in message 1.7.  SS can then decrypt this message to obtain $N_b$, which he returns to B in message 2.7. Hence B believes that A has correctly established a session with him which may led to his arrest.

\textbf{This attack can be described as follows:}

\medskip

1.3 $ 	A 	\rightarrow SS $:	$E(N_a || A : PK_{SS})$

\medskip

2.3 $	SS(A) \rightarrow B $:  	$E(N_a || A : PK_B)$

\medskip

2.6 	$ B	\rightarrow SS(A)$ :    $E(N_a || N_b : PK_A) $

\medskip

1.6 	$SS	\rightarrow A $: 	$E(N_a || N_b : PK_A)$

\medskip

1.7	 $A	\rightarrow SS $ 	$E( N_b : PK_{SS})$

\medskip

2.7  $ SS(A)	 \rightarrow B $ : 	$E( N_b : PK_B)$

\medskip

In the same paper Lowe showed that it is easy to change the protocol so as to prevent the attack; for this purpose he included the responder's identity in message 6 of the protocol.

Hence, the step 2.6 of the attack would become $E(B || N_a || N_b : PK_A)$  and the intruder cannot successfully replay this message in the step 1.6.

\subsubsection{Use of the identities in the authentication protocol}

It is clear that the Needham-Schroeder-Lowe protocol will allow journalists and whistleblowers to authenticate effectively. However, the use of identities in authentication steps can be dangerous for a whistleblower, because if the Tartuffe's administration was able to insert a kleptographic backdoor or if the SS had obtained a clear message from Bob or Alice,  they would be sure to find the identities or pseudonyms of the interlocutors of the journalist or the whistleblower.

For example, in July 1990, in South Africa, the Apartheid regime had managed to find a group of  ANC  activists involved in the fight, due to the fact that some of them did not take care of their files and the discs containing encryption keys \cite{Jenkin}.

 It may also be noted that the use of a kleptographic backdoor such as the supposed Dual\_EC\_DRBG cryptographic pseudo-random number generator would allow the SS to read the content of the messages, $E(B || N_a || N_b : PK_A)$ , which will directly give the identity of the protagonists.

\subsubsection{ Non-identity authentication}

At the time of kleptographic backdoors imposed by gouvernments and mass surveillance, peharps the secret services will be able to read the messages of whistleblowers and journalists, which implies the importance of reducing their risks. 
The first step would be, for example, to remove their identities in the Needham-Schroeder-Lowe protocol. However, removing identities and pseudonyms without replacing them by something else, would put the principal in a more dangerous situation since any attacker, even without a backdoor, could use the Lowe attack \cite{Lowe} against him.

\textbf{The identity replaced by artificial intelligence tools}

Introduced in \cite{Kane}, this paper removes the explicit identities present in the Needham-Schroeder-Lowe which are replaced by partials quotients. The Artificial Intelligence tools will allow the principals to guess what should be the identity of the person who wishes to authenticate.

\textbf{Example of an authentication protocol without an identity}.

Let's suppose that the Alpha Bank has several clients including Alice and Bob and Trudy who wish to authenticate without allowing Tartuffe to know that the accounts belong to them.

The key $ K_1 $ belonging to Alice used to connect every day at around 12am from downton Dakar, in addition, her connection is very slow, it takes around 2 minutes to load the page. 

The key $ K_2 $ belonging to the Bob connects from time to time, around 5 pm, from downtown Sevres. 

The key $ K_3 $ belonging to Trudy connects every day at 6pm from Obelisque Square, he takes the time to verify the data he sends, that is why he types around 15 words per minute.

Alice sends to Alpha Bank at 11:50 from downtown Dakar, a nonce $ N_a $ encrypted with the public key of the bank $ E (N_a: PK_A) $. Given the time, duration, and the location of the connection, Alpha presumes that  it is Alice and prepares his reply.

Alpha computes the 9 partial quotients, noted $ FC_1 $, obtained from the transcendantal number (  $\sqrt[3]{\log(N_a)}$ ) and the hash of the concatenation of the public keys of Alice and the Bank ($ H(K_1| PK_A) $). This hash corresponds to the vector $ Y_1Y_A$ in the paper \cite{Kane}. Alpha chooses a nonce $ N_b$ and returns to Alice $ E (FC_1 || N_b: K_1) $.

Alice verifies the conformity of the partial quotients received and replicates by computing  the nine partial quotients, noted $ FC_2 $, obtained from the transcendental number (  $\sqrt[3]{\log(N_b)}$ )and the hash of the concatenation of the public keys of the Bank and Alice ($ H(PK_A | K_1) $). This hash corresponds to the vector $ Y_AY_1 $ in the paper \cite{Kane}. 
Alice sends to Alpha $ E (FC_2: PK_A) $.

\textbf{Summary of the protocol}

\medskip

$ Alice \rightarrow 	Alpha	$ :    	$E(N_a : PK_A)$

\medskip

$ Alpha \rightarrow Alice	 $ :   	  $E(FC_1||N_b  : K_1)$

\medskip

$ Alice 	\rightarrow Alpha	$ :    	 $E(FC_2  : PK_A)$.

\medskip

Note: Unlike the choice of the paper \cite{Kane}, in our schema we assume that Bob and Alice do not need to use an identity based-cryptosystem, to know their respective keys. Since they had already exchanged their keys or they can guess the owner of a key via IA's tools.

\subsection{ Whistleblower's schemes}

Through Find My app, we have a private key $ d_i $ and a public key $ P_i $ issued every fifteen minutes. We will divert the use of these keys $ d_i, P_i$ to decrypt or encrypt the messages of the whistleblower. Similarly we can use these keys to help in the authentication of the players, or to put in place an appointment, following one of the four models.

\subsubsection{Model 1}

\textbf{ Context:} This is the case where the Whistleblower knew the journalist or activist who will disseminate the confidential information on a broad scale. Here, Alice possesses confidential informations she wishes to transmit to Bob, while knowing that the Internet and cellular networks are subject to mass surveillance. If Bob diffuses the information after receiving an email from Alice, it is likely that the SS could find Alice quite easily. We will denote in the following, the appointment location by $square \, x$.

\textbf{Prerequisite: } Alice and Bob calculate some public keys  ( ${P_i}_{Alice} \dots {P_n}_{Alice}$ and ${P_i}_{Bob} \dots {P_n}_{Bob}$) which they exchange before the start of the operation. Alice and Bob will break all contacts after this prerequisite. 

\textbf{Steps:}

\textbf{Step 1:} As soon as Alice obtained confidential informations on Tartuffe's administration, she sends by using the Find My app of other users, an address for the meeting with Bob via the ICloud $E(place x: {P_i}_{Bob} )$ $ H({P_i}_{Bob})$ ( she just had to broadcast $ {P_i}_{Bob}$ near a user of Find My app).

\textbf{Step 2:} By searching on ICloud every day, Bob uses $ H({P_i}_{bob}) $ to find if there is an appointment location where he can meet with Alice. In this case he will find $place \, x$.

\textbf{Step 3:} At the time of the appointment (it may be the time when the message where sent in ICloud) Bob and Alice will use messages encrypted with their public keys, they can also use the authentication protocol  presented in this section before beginning the conversation. 

\textbf{Summary:} Alice has used Find My app to set up a meeting with Bob without having to send a compromising email.
During the meeting, Alice and Bob may also use Bluetooth paring, however, we do not recommend this method since the code can be easy to attack because of its small size, moreover it can exist some flaws on the Bluetooth protocol \cite{Antonioli} known by the SS.

\medskip

\subsubsection{Model 2}

\textbf{Context:}

 It is the case where the whistleblower did not know the journalists or activist before the beginning, however she may find them during the operation. 
 Alice, who is working for the Tartuffe's administration has obtained confidential information showing that Tartuffe is trying to exterminate migrants by throwing them to the crocodiles and the venomous snakes. To stop this project, she wants to transmit this information to journalists knowing that the Internet and cellular networks are under masse surveillance.

\textbf{Prerequisites:}

 Alice knows that journalists come every day to attend the Tartuffe's press conference,  she decides to use the moment where the journalists turn off their wifi and cellular signal to recover the public keys broadcast by their devices through the application Find My ($ {P_i}_{journalist_1}, \dots, {P_i}_{journalist_n} $ and $ {P_i}_{SS_1}, \dots, {P_i}_{SS_m}, {P_i}_{Tartuffe} $).

We may note that, she will be oblige to recover the keys of her opponents as she could not distinguish them from the key of the journalists.
Knowing that the $ P_i $ broadcast during the Tartuffe's press conference could be used by a administrative whistleblower, journalists will keep their private keys corresponding ($d_i$).
 Alice and Bob have no physical contact or on the Internet during these prerequisites.

\textbf{Steps:}

\textbf{Step 1:} Alice chooses a place where there is a large number of people who are ready to help whistleblowers (University, subway, Coffee Shop, \dots) and recovers the large number of keys $ {P_i}_{anonymuser} $ sent by Broadcast Bluetooth by the users of Find My app. Among these  anonymous users, there are users who have lost their devices, others are activists who want to help whistleblowers, others are Tartuffe's partisans or SS.  
She sends the secret information previously encrypted with the keys of the people present in the press conference by broadcast Bluetooth.  $E(secret: {P_i}_{journalist_1}), \dots, E(secret: {P_i}_{journalist_n})$ and $E(secret: {P_i}_{SS_1}), \dots, E(secret: {P_i}_{SS_m}), E(secret: {P_i}_{Tartuffe})$.

\textbf{ Step 2:} Alice sends to the ICloud, via Fin My app users the place where she would like the journalist to come get the secret informations $E(square \, x: {P_i}_{journalist} )$$ H({P_i}_{journalist})$. She will also do the same thing for all the anonymous users who have to deposit the secret at $square \, x$,
 $E(square \, x: {P_i}_{anonymoususer} )$  $H({P_i}_{anonymoususer}) $.

\textbf{Step 3:} By searching on ICloud Frequently, Bob uses $ H({P_i}_{journalist}) $ to find the place where he must go to find the confidential information. 

\textbf{Step 4:} By searching on ICloud, the person who whishes to help the whistleblower, the anonymous user finds the  place where he have to deposit the encrypted confidential information by using $ H({P_i}_{anonymoususer})$ and  $E(square \, x: {P_i}_{anonymoususer} )$.  

\textbf{Step 5:} At the time of the appointment, the anonymous user sends in Broadcast Bluetooth the encrypted information received from Alice. The journalists and the SS will use their private keys $ d_i$ to find the confidential information.

\textbf{Summary:}
 Alice used the advantages given by Find My app to encrypt confidential information, to set up a meeting between  journalists and the anonymous users without having to send compromising email, or to meet journalists physically which would allow the SS to find her.

\medskip

\subsubsection{Model 3}

\textbf{Context:}

It is the case where the whislbleblower does not trust journalists and activists, she whishes to broadcast the confidential information directly to the population. 
Alice discovers that Tartuffe is a dangerous pathological liar who puts the children of migrants in cages to torture them. She decides therefore to disseminate that information without passing through the journalists or the activists. However, sending encrypted confidential information on social networks may be dangerous due to the mass surveillance program. Similary sending by Bluetooth without encryption on the street could also be dangerous since the SS or Pongo could recognized her.

\textbf{Prerequisites:}

 Alice must choose locations where she could find help more easily (University, Coffee shop, \dots ), for example she will need people ready to go to an appointment, somewhere in the world, to receive the information that Tartuffe's government is hiding to the people.
 Alice will have no physical or virtual contact with those who will receive the secret information. 

\textbf{Steps:}

\textbf{Step1:} by passing near a device broadcasting ${P_i}_{anonymoususer_1}$ through Find My app, Alice save that key and encrypt the secret information with that latter ($E(secret: {P_i}_{anonymoususer_1} )$).
Later when Alice will be in a crowd, she will send by broadcast Bluetooth, the encrypted secret and the public key of the  anonymous $ user_1$ ($E(secret: {P_i}_{anonymoususer_1} )||{P_i}_{anonymoususer_1}$) . 

\textbf{Step 2:} whenever Alice saves other keys broadcasted by Find My app, for example, ${P_i}_{anonymoususer_2}$, she will repeat the previous operation. Hence, she will broadcast this information as much as possible.

\textbf{Step 3:} the people present and who wish to help the Whistleblowers will set up an appointment to the owner of the key $ {P_i}_ {anonymousser_1} $, by sending to the ICloud the location chosen for the meeting ($ square \, x $ chosen will vary depending on the helper), accompanied by the hash of the key received ($E(square \, x: {P_i}_{anonymoususer_1} )$     $H({P_i}_{anonymoususer_1})$.

\textbf{Step 4:} by looking for his lost device,   $anonymoususer_1$ will calculate the hash of his public keys broadcast by Find My app, and verify on ICloud ($ H({P_i}_{anonymoususer_1})$) where his device is located. He will see some places where his device is located, he will also see locations where his devices can not be found, which means that there are some potential appointment meetings on these locations.

\textbf{Step 5:} at the appointment time, the helper who had received the package  $E(secret: {P_i}_{anonymoususer_1} )||{P_i}_{anonymoususer_1}$ will send by broadcast Bluetooth $E(secret: {P_i}_{anonymoususer_1} )$ allowing the anonymous user to find secret sent by Alice.

\textbf{Summary:}

 Alice used Find My app to find people ready to help her disseminate the confidential information very widely. People who wanted to help  will set up meetings by using Find My app, allowing Alice to disseminate the secret without having to send emails or  to be present at the time of the broadcast of the confidential information.

\medskip

\subsubsection{Model 4}

\textbf{Context:}

This is the case where the whistleblower knows the place and time of the meeting but she ignores the identity of the journalist. Activists and journalists suspect that Tartuffe is obliging some illegal immigrants to work in his residences without paying them, but for lack of evidence and knowing that these illegal workers will be afraid to testify, or to contact them on the social networks -due to mass surveillance-, the journalists will block that information.  A person of goodwill will set up a meeting (location and time) for Alice and Bob. The meeting's location varies very lightly because Alice and the journalist should not to be seen together, but should be able to communicate with Bluetooth.

Sending Broadcast Bluetooth messages without encryption could also be dangerous for Alice, since Pongo could be, by chance present allowing him to intercept the message and to recognize the Tartuffe's employee.

\textbf{Prerequisites:}

We will assume that the time and place of the appointment remain secret for all, even the SS.

 The person of goodwill must choose a place well covered by WiFi and cellular which will allow to distinguish the signal of the journalist and Alice. 

The distance between Alice and the journalist is equal to the effective range of a reliable Bluetooth. We will notice in the following, the place chosen for Alice by $x $.

We assumed that the appointment locations are only known by three protagonists (Alice, Bob and the person who set up the meeting). The two public keys  (${P_i}_{Alice}, {P_i}_{Bob}$) must be exchanged very quickly, and if a third party put a third key, this last one would not be taken into account by Alice and Bob.

 It is to prevent the irruption of an SS that the locations and the time of the meeting are kept secrets. In addition, the two keys exchanged quickly makes it possible to avoid the irruption of a Pongo present by chance on the premises at the agreed time.

\medskip

\textbf{Steps:}

\textbf{ Step 1:} Presents a few minutes before the appointment's time at the meeting's location, Alice and the journalist ensure that there is no device using Find My app in broadcast Bluetooth in this place.

\textbf{Step 2:} At the time of the appointment, Alice turn off her WiFi and cellular allowing her Find My app to send by Bluetooth ${P_i}_{Alice}$.

\textbf{Step 3:} The journalist recovers ${P_i}_{Alice}$ through Find My app, other people presents may also receive the public key of Alice.

\textbf{Step 4:} Bob turn off his wifi and cellular, allowing Alice to recover his public key  ${P_i}_{Bob}$. The other people presents may receive the Bob's public key, however, they will not be able to distinguish it from Alice's public key (the key change frequently).

\textbf{Step 5:} Alice and Bob may use their public and private key corresponding to communicate.

\textbf{Summary:}

By diverting the purpose of Find My app, Alice and Bob were able to authenticate and to create a secured channel allowing them to discuss without any risk of interception. 
Thus, Alice illegal immigrant and Bob investigative reporter was able to conduct an interview without the identity of one being revealed to the other, which allows the journalist to protect his sources \cite{sources}, since he does not known her, and Alice to express herself without fear since the journalist ignores her face and her name.

\subsection {Security and Efficiency analysis}

The security of the four models presented in this section is based on the security given in native by Apple's applications and on those we have introduced in the design of these models. Thus in this section, we will analyze the securities issues introduced by the Find My app and those induced by our models.

\subsubsection{Threat Model}

As supposed in \cite{Roth}, we also consider in our scheme that the primary security objective of these tools is to conceal the presence of whistleblowers, and to eliminate network traces that may make one suspect more likely than another in a search for the whistleblower. 
We consider in this scheme that the security of the whistleblower is more important than the disclosure of the information.

\subsubsection{Threats in our scope}

In this scheme, a government may threaten Internet providers to have their help in targeted surveillance or to stop their service. In addition, a mass surveillance may also be done by the government and the adversaries may also be able to create a Bluetooth botnet or introduce a backdoor on the public key used by Alice.
We believe that the secret services could try to monitor all messages sent through the Internet and cellular networks. Thus, any encrypted message, issued by a government employee could be followed (tagged). The attacker can also record messages received and sent by government employees.

\subsubsection{Threats not in our scope}

Malware and spyware are not in our scope, the whistleblower and the journalist has to secure their device before any action.

\subsubsection{Security-related assumptions}

We assume that the court does not per se consider that relaying encrypted files as criminal. 
We also suppose that the cryptosystem used possesses the property of indistinguishability.
.

\subsubsection{Security properties}
In the models presented in this paper, there are attacks common to four models and others are specific to each model.

\subsubsection{Commons Attacks }

\textbf{ Kleptographic Backdoor}

Find My app sends by Broadcast Bluetooth the public keys $ P_i $, the attack possible on these public keys would be to introduce on it  a Kleptographic backdoor. This attack would allow the attacking entity to monitor Alice Remotely. Without being able to prove the inexistence of this attack, we can said with a strong conviction that it would be difficult to realize since the choice of parameters by Apple seems good.

\medskip

\textbf{ A Forensic Investigation}

Tartuffe could seize all devices belonging to his employees in order to conduct a forensic investigation, the purpose would be to find the device that sent or encrypted the confidential information. This attack should fail if Alice had encrypted her device or erased the targeted data.

\medskip

\textbf{ Tartuffe could try to forbid the dissemination of this information by cutting the Internet and cellular signal:}

Tartuffe could cut the Internet of the country as it has been done in other countries. However this approach would fail, because he could not cut the Internet during a very large period of time \cite{Shutdown}, in addition a large part of our communications are done by Broadcast Bluetooth and the dissemination can be done in another country or continent.

\subsubsection{Specifics Attacks}

\medskip

\textbf{Attack on model 1:}

\medskip

\textbf{Attack on Alice's anonymity}

\medskip

\textbf{Step1:} During the recovery of the keys

\medskip

Tartuffe, his supporters and the SS can not guess that Alice has infiltrate them in order to  transmit confidential informations to Bob.

\medskip

\textbf{Step 2:} During the delivery of the messages

\medskip

The anonymity of Alice is not threatened, because she set up Bob's appointment by using the Find My app of passer-by in the street, who will relay the appointment to Bob via ICloud. She can also send the appointment to Bob via the Find My app of her device due to the fact that  Apple is supposed to protect the helper's anonymity. In addition, she broadcast encrypted informations and she can be very far from Bob (up to 400m for some devices).

\medskip

\textbf{Step 3:} Physical surveillance to remove the anonymity

\medskip

Even if Tartuffe decided to organize a physical surveillance  of all those who work in his administration, the anonymity of Alice would not be in danger, since she is undetectable when she exchanges keys with Bob. 
Also, the SS will not be abe to prove that Bob and Alice have discussed via encrypted messages sent by Broadcast as their communications can come from any person within a radius of  400m.

\medskip

\textbf{Step 4:} The use of a kleptographic backdoor to lift the anonymat

\medskip

If we assume that Tartuffe has been able to install a backdoor on all the keys broadcasted by Find My app, he could not lift Alice's anonymity of Alice. However, the SS or the Pongo can intercept the messages exchanged between Alice and Bob and if the backdoor system put in place allows to know the owner of each intercepted encryption he may know, but a such system to our knowledge does not exist yet since the cipher is indistinguishable.
 
We can also suppose that President Tartuffe has given to some of his supporters the private key of the backdoor in order to help him fight against the fake news, we will name the partners of the president $ Pongo_1, \dots, Pongo_{43}$.
$Pongo_1$ would intercept by chance on the street $E(N_a : {P_i}_{Bob})$,  $E(FC_1||N_b  : {P_i}_{Alice})$ and $E(FC_2  : {P_i}_{Bob})$. That will allow him to see $N_a, N_b, FC_1, FC_2$, however if Alice and Bob had used Needham-Schroeder-Lowe, $Pongo_1$ would see $N_a, N_b, Alice, Bob$, which would lifted the anonymity of Alice and Bob.

\medskip

\textbf{Prevent the exit or the delivery of confidential informations}

\medskip

 Tartuffe could try to avoid the dissemination of the confidential information by taking a number of attacks against this model: 

\medskip

\textbf{Step 1:} Forbid the the exchange of the keys

\medskip

This would fail, since Tartuffe, the SS and the Pongo are not present at the time of the exchange of the keys.

\medskip

\textbf{Step 2:} To set up fake meetings

\medskip

In order to avoid a trap, an authentication on ICloud when setting up the meeting may be necessary in some cases. For example, if Pongo manages to steal Alice's public key, he would be able set up a meeting with Alice by sending to the ICloud $E(square \, x: {P_i}_{Alice} )$  $ H({P_i}_{Alice})$.
With this action, Pongo may perhaps identify and arrest Alice. 
However, if Alice and Bob authenticate on ICloud through Find My app and by using the authentication algorithm of this section, Pongo could not attract Alice in a mousetrap.

\medskip

\textbf{Summary of the protection against the trap:}

\medskip

Instead of sending the appointment location directly on ICloud, Alice sends a nonce $ N_a$ chosen randomly, on the ICloud. $H({P_i}_{Bob})$ is the index which will allow Bob to retreive $N_a$.

\medskip

$Alice 	\rightarrow Bob $	:    	$E(N_a : {P_i}_{Bob})$ $H({P_i}_{Bob})$

\medskip

Bob reply on ICloud by sending $N_b$ and the partial quotients $ FC_1$, calculated from $ N_a$ and hash of the contatenation of the public keys of Alice and Bob. $H({P_i}_{Alice})$ is the index which will allow Alice to retreive $FC_1, N_b$.

\medskip
 
$Bob 	\rightarrow Alice $	:   	  $E(FC_1||N_b  : {P_i}_{Alice})$  $H({P_i}_{Alice})$

\medskip

Alice computes and transmits via the ICloud the partial quotients, $FC_2$, calculated from $ N_b $ and the hash of the concatenation of the public keys of Bob and Alice.

$Alice 	\rightarrow Bob $	:    	 $E(FC_2  : {P_i}_{Bob})$ $H({P_i}_{Bob})$.

\medskip

\textbf{Attacks on model 2: }

\medskip

\textbf{Attack on the anonymity of Alice}

\medskip

\textbf{Step 1:} During the press conference

Although working in the administration of Tartuffe, Alice can collect the public keys without being spotted by the SS. We recall that the public keys $ P_i$ are sent in broadcast Bluetooth to users of Find My app.

\medskip

\textbf{Step 2:} During the delivery of the messages:

At no time Alice's anonymity can be lifted since she sends encrypted information and she can be very far from those receiving this message (up to 1000m for some devices).

In order to create more confusion to the SS and the Pongo, a person of goodwill can relay the message in other places, that is to say, recover the keys from others anonymous users and set up a meeting with journalists (like Alice did it).

Arresting the anonymous user who deposits the secret to the journalist would bring no information on Alice's identity, since he does not know  Alice, nor the journalist, nor the content of the message. In addition it will be difficult for the SS to identify the person of goodwill since he is anonymous and at the meeting, this person can be very distant from journalists.

\medskip

\textbf{Step 3:} Physical surveillance against employees to remove the anonymat

Even if Tartuffe organizes a physical surveillance of all those who work in his administration, the anonymity of Alice would not be in danger since she is undetectable when she recover the  keys  and she can not be accused of sending the Bluetooth since the sending can come from any person in a radius of 1000m, so legally, it would be impossible to accuse Alice.

\medskip

\textbf{Step 4:} The use of a kleptographic backdoor to lift the anonymity

Even if we assume that Tartuffe has been able to install a backdoor on all the keys broadcasted by Find My app, he could not lift the anonymity of Alice since she encrypts with the keys of people present at the press conference (journalists, SS, Tartuffe) and she organizes meetings with the keys of anonymous users.

\medskip

\textbf{Forbid the output or delivery of the confidential information}

  Tartuffe could try to forbid the dissemination of this information by trying a number of attacks: 

\medskip

\textbf{Step 1:} Forbid the key's delivery 

Scrambling the Bluetooth signal will forbid Alice to recover the public keys, however, in addition to the trouble that this interference will cause to his administration, Alice could circumvent this censorship by taking the keys in places frequented by journalists and activists (universities, subway station, coffee shop  \dots). 

\medskip

\textbf{Step 2:} Organize fake appointments

To discourage journalists present at the press conference to go pick up the confidential information sent by whistleblowers, Tartuffe's SS could set up fictitious appointments. Their presences in these appointments would be a waste of time, since there would be no information available. The journalists can avoid this attack by working in pool, and asking good willing people to send the encrypted confidential information directly to the journalists' pool.

\medskip

\textbf{Attack on model 3:}

\medskip

\textbf{Attack on the anonymity of Alice}

\medskip

\textbf{Step 1:} During the process of recovering the keys

Although working in Tartuffe's administration, Alice can receive the public keys $ P_i$ on the street without being suspected by the SS, since the purpose of Find My app is to recover the public keys sent by other Apple's devices in Broadcast Bluetooth. 

\textbf{ Step 2:} During the delivery of the message:

Alice encrypts the confidential informations with an anonymous public key found in the street, and she sends the cipher in broadcast mode to groups of anonymous people who ignore her identity. In addition, Alice may send the cipher far from the anonymous people, hence, Alice's anonymity will be difficult to lift.

In order to create more confusion to the SS and the Pongo, good wills can relay the message in other places, that is to say, send the package received to others by broadcast.

Arresting the people who carry the encrypted secret would bring no information on Alice's identity, since they know nothing on Alice, nor the anonymous person to whom the public key $ P_i $ belongs.

It will be very difficult for Tartuffe and his SS to identify the people who exchange the secret in the street since they are anonymous and these persons can be very far one from another  and they are not under the mass surveillance (internet, cellular, \dots). However the video surveillance may be a threat in some case.

\textbf{Step 3:} Physical surveillance to remove the anonymity

Even if Tartuffe decides to organize a physical surveillance of all those who works in his administration, the anonymity of Alice would not be in danger since she is undectable when she receives the keys of the anonymous people in the street and she can not be accused of sending it by Bluetooth, since the sending can come from any person present in a large radius. So legally, it would be difficult to accuse Alice.

\textbf{Step 4:} Using a kleptographic  backdoor to lift the anonymity

Even if we assume that Tartuffe has been able to install a backdoor on all the keys broadcasted by Find My app, he could not lift the anonymity of Alice since she encrypts with the keys of anonymous people. The meetings are also scheduled with keys that do not belong to Alice.

\textbf{Forbid the delivery of confidential information}

Tartuffe might try to complicate or forbid the dissemination of this confidential information by taking some attacks:

\textbf{Step 1:} Forbid the key's broadcasting

By scrambling the Bluetooth signal throughout the city or prohibiting the use of Find My app, Tartuffe can jeopardize Alice's keys recovery, however, this prohibition would be very difficult to implement, since many hospitals and other administrations are using Bluetooth, and this latter does not depend on an Internet Provider which needs a licence given by the Tartuffe's administration.

\textbf{Step 2:} Organize fake appointments

To discourage a person of goodwill to go pick up the confidential information sent by whistleblowers, Tartuffe's SS could set up fictitious appointments. The presence at these appointments would be a waste of time, since there would be no information available. 

The owner of this confidential information can avoid this attack by sending in broadcast Bluetooth the $ d_i $ (corresponding to $ P_i $) accompanied by the time and location of the appointment received on ICloud. Hence, the person of goodwill will not lose his time if it is a fake appointment and those who receive it, will relay it, until it reaches those who are in the area of the appointment, who will recover the secret information and copies it to others.  

\medskip

\textbf{Attack on model 4:}

\medskip

\textbf{ Attack on the anonymity of Alice}

\medskip

\textbf{Step 1:} During the process of recovering the keys

Although working for Tartuffe, Alice can recover the public key $ P_i $ on the street without being suspected by the SS since the appointment with the journalist is confidential. 

\textbf{Step 2:} During the delivery of the message

Alice and Bob discuss with anonymous public keys exchanged on the street, which implies the difficulty of lifting their anonymities.
In addition, it will be very difficult for Tartuffe and his SS to identify Alice and Bob who discuss in a crowded place via encrypted  Broadcast Bluetooth, unlike the Internet where the SS can establish a mass surveillance.

\textbf{Step 3:} Physical surveillance to remove the anonymity

 Even if Tartuffe organize a physical surveillance of all those who work in his house, Alice does not risk to be lynched since she is undetectable during the keys exchange and at no time she physically meets the journalist.

\textbf{Step 4:} Using a kleptographic  backdoor to lift the anonymity:

Even if we assume that Tartuffe has been able to install a backdoor on all the keys broadcasted by Find My app, he could not breach Alice's anonymity, since the cipher is indistinguishable. However, the message content may be accessible to Tartuffe's SS if they are present, by chance, in the premises and time of appointment, which would be very difficult to realize.

\textbf{Forbid the delivery of confidential information}

Tartuffe might try to complicate or forbid the dissemination of this confidential information by taking some attacks:

\textbf{Forbid the key's broadcasting:}

By prohibiting the use of Find My app in his house, Tartuffe can jeopardize Alice's keys recovery, however, she can borrow another device.

\textbf{Organize fake appointments:}

This attack does not apply in this model since the appointement is set up by a trustworthy person.

\subsection{Others advantages }

Find My app could also allow those who have Apple's devices to help someone who is looking for missing or dangerous people.

\medskip

\textbf{Context:} Bob is looking for his missing relative and he knows that he can not trust the authorities \cite{Mexico}, then he looks for help on the social networks. 

\medskip

\textbf{Step 1:} Alice downloads a facial recognition software.

\medskip

\textbf{Step 2:} Alice downloads the picture and the public key of Bob who is looking for the missing person ($ {P_i}_{search} $).

\medskip

\textbf{Step 3:}  Alice's software activates the camera of her device, allowing her to compare the faces of people encountered in the street, in the subway, \dots with those of the missing person.

\medskip

\textbf{Step 4:} As soon as Alice meets the missing person at ($ square \, x $), her device sends to the ICloud via Find My app $H({P_i}_{search})$ and $E(square \, x: {P_i}_{search})$, which will allow Bob to find his missing relative without the help of the authorities. 

\medskip

\textbf{Remarks:} 

\medskip

The use of this process should be monitored by users to avoid that software being used to track  political opponents or whistleblowers.

\section{Inconvenients introduced by Find My app}

Beside the advantages, the introduction of the Find My app may weaken or endanger the safety of whistleblowers in some cases. Many questions and worries were raised on some possibles securities issues \cite{{critics},{wired}}, here, without being exhaustive, we will in this section describe two problems  introduced by this application. 

\subsection{Model 1: Using Find My app to track journalists and whistleblowers}

\textbf{Context:}

 Tartuffe knows that whistleblowers and journalists meet somewhere in one place that he ignores and in one hour which he suspects.

\medskip

\textbf{Steps}

\medskip

\textbf{Step 1:} Tartuffe's SS interrupt the Internet signal during this time period, allowing them to intercept the $P_i$'s broadcasted by Apple devices.

\medskip

\textbf{Step 2:} A botnet of  Tartuffe's partisans ($Pongo_1, \dots, Pongo_n$) will recuperate during this period of  time the $P_i$'s broadcasted by all Apple's device anywhere in the city.

\medskip

\textbf{Step 3:} The SS collect the $P_i$ broadcasted by whistleblowers and journalists by targeting their houses, offices, headquarters or street demonstration.

\medskip

\textbf{Step 4:} The SS compare the keys recovered by the Pongo with the keys they captured. 

\medskip

\textbf{Step 5:} If the key captured near the Alice's home  ${P_i}_x$ is found next to the supposed key of Bob $ {P_i}_y $ somewhere in the city, then Tartuffe would conclude that Alice is the traitor of his administration who informs Bob. Similarly, the keys found among the protesters would allow the regime to target their home.

\medskip

\textbf{Remark:}

Apple's devices used by one person should not broadcast the same key, for example they may add random numbers on each device, hence when the IPhone broadcasts $P_i + a_1$, the IPad of the same person should broadcast $P_i + b_1$. Knowing that $a_1$ and $b_1$ are two differents random numbers.

\subsection{Model 2: Using Find My app to find and ban a religious minority in the country.} 

\textbf{Context:}

Tartuffe decides to ban all members of a religious group who live in his country with the help of the SS and the Pongo. 

\medskip

\textbf{Steps:}

\medskip

\textbf{Step 1:} Tartuffe's supporters use a facial recognition software to track people from this religious group (ethnic origin, facial shape, clothing).

\medskip

\textbf{Step 2:} As soon as the facial recognition supposes that someone is a member of that religious group, their application would send to the ICloud $H({P_i}_{Aliens})$ et $E(square \, x: {P_i}_{Aliens})$, where ${P_i}_{Aliens}$ is the key that the SS have made available to Tartuffe's supporters to help them find the undesirables. 

\medskip

\textbf{Step 3:} By searching in ICloud, the SS will quickly find the $ Squarre \, x$ where they can arrest and deport the suspected member.

\medskip

\textbf{Remark:}

This religious profiling can in some cases be accompanied by a remuneration for those who would agree to participate in this tracking. This tracking is dangerous and must be prohibited by laws. 
In addition to these attacks, others can use the Find My app to track whistleblowers as using Trojan to collect the $P_i$'s. Thus, the limits of these attacks depends on the creativity of their designers.

\section{Conclusion}
In this paper, we had diverted a little addition of Apple, intended to find lost or stolen devices, in a powerful tool for whistleblowers, in order to get out of confidential informations while keeping their anonymities. 

We also tried, to put in place a type of broadcasting that would protect the identity of the whistleblower even in a world targeted by mass surveillance.

Finally, even if the government was able to impose the use of backdoors, the use of the schemes presented here should allow the whistleblowers to remain anonymous. 

It should be noted that the inconvenients introduced by Find My app are serious, and should be corrected technically and legally, however these corrections should not harm  the many benefits that this application could bring to whistleblowers.

The limit of the schemes using Find My app depends on the creatively of journalists and whistleblowers, hence, it would be interesting in the future to design a scheme using the public keys broadcasted by Find My app and the blockchain  technologies to fight deepfakes.


\end{document}